\documentclass[reprint, amsmath, amssymb,aps, pre,nofootinbib]{revtex4-1}
\usepackage{graphicx}

\usepackage[colorlinks, linkcolor=blue, citecolor=blue, urlcolor=blue]{hyperref}

\begin{document}

\title{Transient dynamics in the thermal ratchets transport model}
\author{Abhranil Das}
\thanks{Present address: The University of Texas at Austin, Physics Department, C1600 (RLM), Austin, Texas 78712, USA}
\email{abhranil.das@utexas.edu}
\author{Soumitro Banerjee}
\email{soumitro@iiserkol.ac.in}
\affiliation{Indian Institute of Science Education and Research, Kolkata\\
Mohanpur Campus, Nadia 741246, West Bengal, India}
\date{\today}

\begin{abstract}
The thermal ratchets model toggles a spatially periodic asymmetric
potential to rectify random walks and achieve transport of diffusing particles. We numerically solve the governing equation for the full dynamics of an infinite 1D ratchet model in response
to periodic switching. Transient aperiodic behavior is observed that converges asymptotically to the period of the switching. We study measures of the transport rate, the transient lifetime, and an equivalent of `amplitude', then investigate their dependence on various properties of the system, along with other features of the transient and asymptotic dynamics.
\end{abstract}

\maketitle

\section{Introduction}

\begin{figure}
\includegraphics[width=1\linewidth]{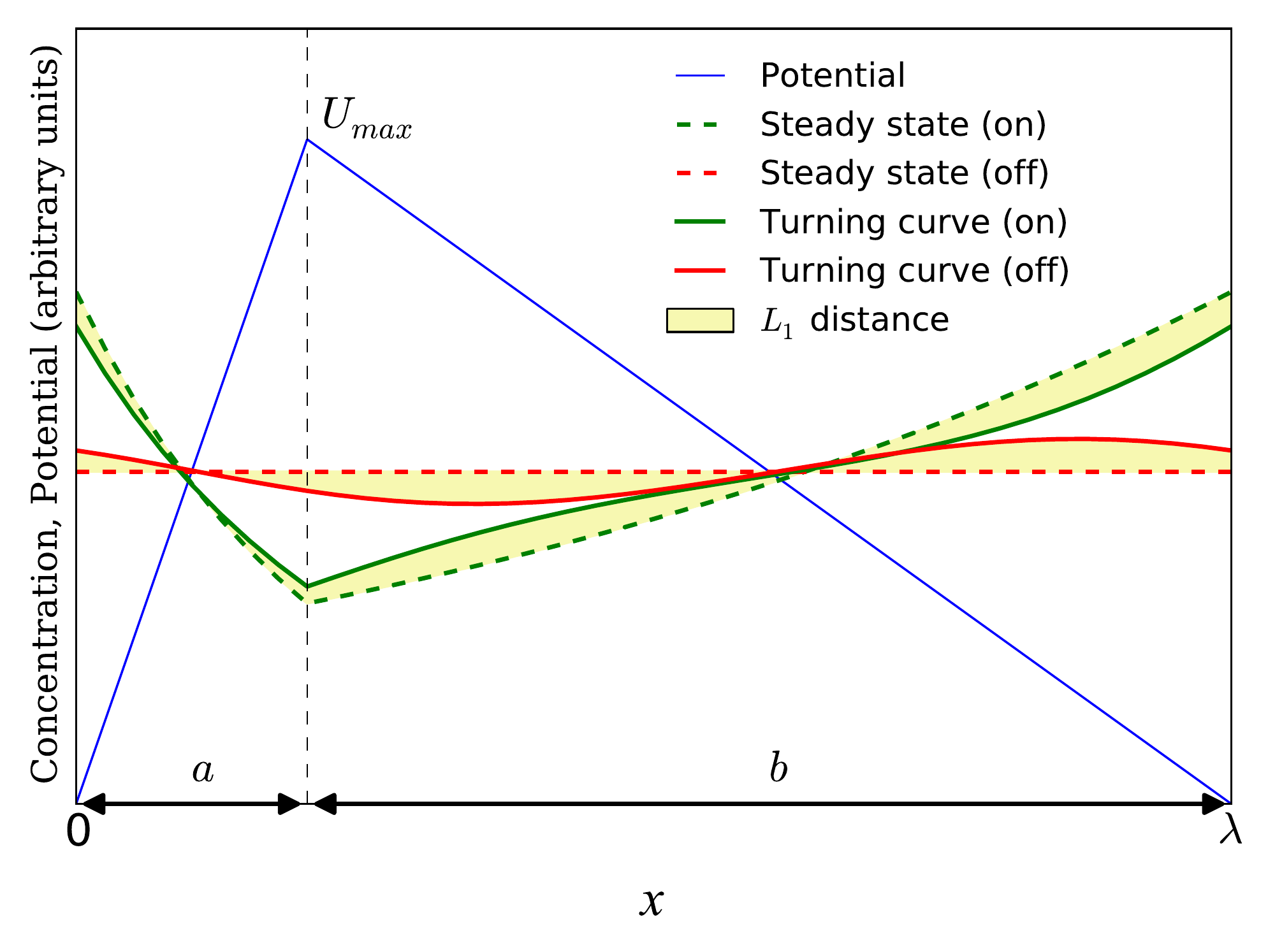}
\caption{(Color online) A period of the ratchet potential. Dashed curves are analytically
obtained equilibrium Boltzmann distributions in response to it, in
the \textsc{on} and \textsc{off} states. Green and red solid curves,
obtained from the numerical solution, are the `turning curves' of
the concentration at the end of \textsc{on} and \textsc{off} phases
respectively, once periodicity has been attained. The filled area
corresponds to the $L_{1}$-distance discussed and plotted in Fig.
\ref{fig:L1vsperiod}. Parameters are $\alpha=\frac{1}{5},\hat{D}\approx.05,\hat{U}_{\text{max}}\approx.04.$\label{fig:turning-curves}}
\end{figure}

Pure diffusion is isotropic, not resulting in transport in any particular direction. A potential gradient induces drift in addition to the diffusion, and may result in transport. In overdamped environments, any external force induces a proportional drift velocity $v=F/\gamma=-\frac{1}{\gamma}U_{x}$ (subscripts denote partial derivatives),
where $U$ is the potential energy. Thus, a potential whose space-averaged
force is zero will not naively be expected to result in non-zero average transport. But the thermal ratchets model achieves this by rectifying diffusion \cite{1997Astumian,2002Reimann}.
The most common form studied is a spatially periodic asymmetric sawtooth
(`ratchet'-shaped) potential (see Fig. \ref{fig:turning-curves}),
exerted over a 1D array of diffusive particles. The potential is switched
\textsc{on} and \textsc{off} either periodically or stochastically,
and results in the transport of diffusing particles in the direction
of the gentler sawtooth slope.

Possible applications of the model
are in approximating transport phenomena by molecular motors \cite{1990ValeOosawa,1994Peskin},
or to construct experimental or theoretical setups that achieve long-range
transport of diffusing particles without requiring a long-range potential
gradient. Several previous works have analyzed the model via either an analytical Fokker-Planck approach, Monte Carlo
simulations, or experimental setups, and have investigated in particular the resulting particle flux and its dependence on various factors \cite{1992DoeringGadoua,1993Magnasco,1994Astumian,1994Prost,1994Rousselet,1998Kamegawa}.

The governing equations for the particle concentration cannot be solved analytically for the full dynamics in general, only under certain assumptions (say spatio-temporal periodicity), or to yield the steady state. To study
the full dynamics, we employ a finite
difference numerical method, with the potential
switched at equal \textsc{on} and \textsc{off} periods. Our results
show that the dynamics is initially aperiodic in time.
With continuing switching cycles, it asymptotes to the period of the switching, during which the concentration oscillates between shapes we shall call `turning curves'. We shall study the dynamics of the particle concentration and investigate the transport rate for the transient and the asymptotic periodic behavior. We also investigate the dependence of these on various system parameters.

\section{Analysis}

In an infinite 1D system of independent particles, the dynamics of
their concentration is governed by the following Fokker-Planck equations
which are the drift-diffusion equations:

\begin{eqnarray}
J & = & vc-Dc_{x}\label{eq:J}\\
c_{t} & = & -J_{x}=Dc_{xx}-(vc)_{x}=Dc_{xx}+\frac{1}{\gamma}(cU_{x})_{x}\label{eq:cdot}
\end{eqnarray}

Here $v$, $c$, and $J$ are the spatio-temporal profiles of the
drift velocity, concentration and flux respectively. In this paper
we consider a constant diffusivity $D$.

Let us adjust for the scales of the system. Say the ratchet potential
has slope-widths $a$ and $b$, spatial period of $\lambda=a+b$ (see Fig. \ref{fig:turning-curves})
and temporal period of $\tau$ (i.e., switched \textsc{on} and \textsc{off}
at intervals of $\tau/2$). Then we can take $D_{0}=\lambda^{2}/\tau$
and $E_{0}=\gamma D_{0}$ as scales for diffusion coefficient and
energy in the system respectively, and using these, (\ref{eq:cdot})
can be recast into the dimensionless form:

\[
\hat{c}_{\hat{t}}=\hat{D}\hat{c}_{\hat{x}\hat{x}}+(\hat{c}\hat{U}_{\hat{x}})_{\hat{x}},
\]

where $\hat{t}=t/\tau$ and $\hat{x}=x/\lambda$ measure time and
space in multiples of the temporal and spatial ratchet periods, $\hat{c}=\lambda c$
is a dimensionless concentration density, $\hat{D}=D/D_{0}$ is a
scaled diffusion coefficient, and $\hat{U}=U/E_{0}$ is a scaled potential
energy. Defining the form of the asymmetric sawtooth potential requires
specifying the degree of skewness of the ratchet shape $\alpha=a/\lambda$,
and the scaled amplitude $\hat{U}_{\text{max}}=U_{\text{max}}/E_{0}$
(see Fig. \ref{fig:turning-curves}). These and $\hat{D}$ constitute
the three dimensionless parameters that govern the character of our
system. Results that follow shall
be accompanied by the parameter values at which particular behavioral
features are highlighted. However, we shall also discuss dependencies on dimensional parameters such as $D$ and $\tau$, as their variations are more practical to consider in a physical setup.\\

Solved analytically for the steady state, the particle density lies
in a Boltzmann distribution in response to the potential profile.
By Eq. (\ref{eq:cdot}), the steady state condition $c_{t}=0$ enforces
uniform flux over space. It may be shown that the flux in fact vanishes
everywhere \cite{MSThesis}, unless there exists an additional external
force, in which case there is a uniform flux in its direction. However,
switching the potential at finite intervals prevents the attainment
of the steady state, but achieves transport even in the
absence of any further external force.

At this point we make the simplifying assumption of spatial periodicity with the following argument. Initializing with
a concentration that has the spatial period of the potential, its
ensuing dynamics shall also be periodic at all times, since all terms
of Eq. (\ref{eq:cdot}) and boundary conditions stay identical in each
spatial period. We thus start from a flat (trivially periodic)
concentration and solve the equation on a single spatial period with
periodic boundary conditions, reproducing the behavior expected on
an infinite line.

Inconveniently, the flux $J$ evaluated at some arbitrary position does
not represent the macroscopic transport of particles. Local fluxes may
be countered by opposite fluxes at other positions and therefore are
not a quantifier of the long-range transport. Spatially averaging $J$, however, serves to cancel out internal flows and
yields only the global transport that is the goal of the system.

Now, since the dynamics maintains the spatial period of the potential,
so does the flux. Averaging it over a single period $\lambda$
then serves as well as over all space, and we define the space-averaged
flux, or \textit{transport rate} as (see \cite{2002Reimann}):

\begin{equation}
T(t)=\frac{1}{\lambda}\int_{\lambda}J\:dx=\frac{1}{\lambda}\int_{\lambda}(vc-Dc_{x})\:dx.\label{eq:transport}
\end{equation}

Owing to the spatial periodicity of $c$, the second
term integrates to zero, and we have:

\[
T(t)=\frac{1}{\lambda}\int_{\lambda}vc\:dx=-\frac{1}{\gamma\lambda}\int_{\lambda}cU_{x}\:dx.
\]

Observe now that the transport will vanish whenever the concentration
is a function of the potential, i.e., $c(x)=c(U(x))$, owing to the
periodicity of $U$. Graphically, this is the case when the two pieces of the
concentration over the shallow and steep ratchet slopes are mirrored
and horizontally scaled versions of each other proportional to their
slope widths. The drift speeds on the other hand being inversely proportional
to the slope widths and in opposite directions ensures the net zero
integral. The steady
state Boltzmann distribution $c(x)\propto e^{-U(x)/\gamma D}$ is a special case of this situation.

\begin{figure*}[t]
\includegraphics[width=1\linewidth]{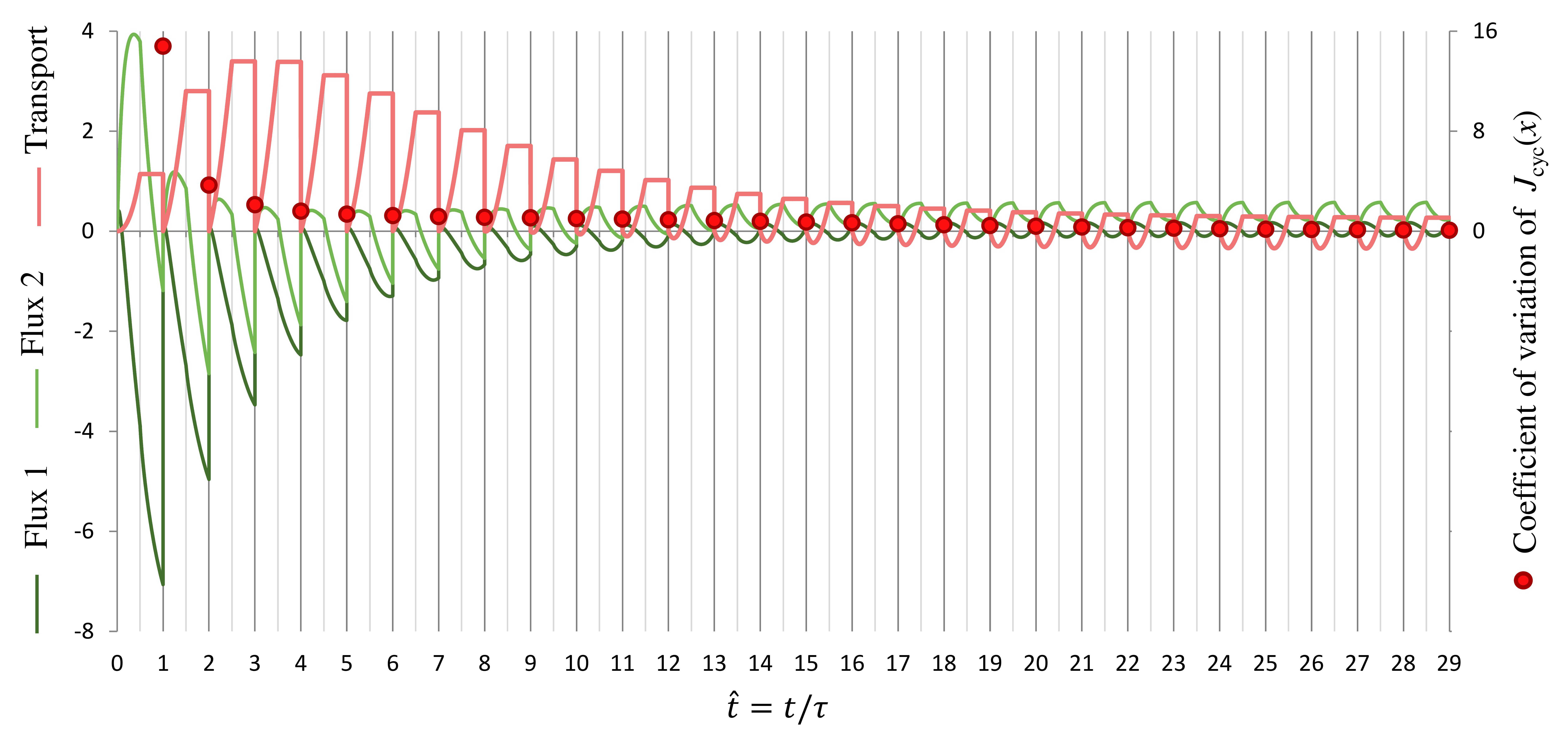}
\caption{(Color online) Cumulative local fluxes through two arbitrary positions and the transport,
normalized as a fraction of the starting (uniform) number density
of particles, against time (measured in multiples of the switching
cycle period $\tau$, so that numbers along the horizontal axis mark
out cycles, and grid-lines at twice the frequency are switching times). The spatial coefficient of variation of the net local fluxes over
a cycle is plotted on the secondary axis. The dynamics converges to
a periodicity after passing through an intermediate maximum. Parameters
are $\alpha=\frac{1}{4}, \hat{D}\approx.006, \hat{U}_{\text{max}}\approx.06.$\label{fig:Cumulative-flux-through}}
\end{figure*}

Zero transport is thus a more general situation than steady state. Even out of equilibrium, there is no transport when $v=0$,
i.e., in absence of drift. This has the following physical interpretation. Pure diffusion from any initial concentration profile relaxes it to a flat one. The local fluxes $J$ of these equilibriating flows
shall depend on the particular initial distribution, but their space-average
is zero, i.e., they achieve no transport. One way to understand this
is in the context of the Green's function: recall that
the particles are independent, hence diffusion from any distribution
is a superposition of diffusion from point sources,
each of which is a spreading symmetric Gaussian with zero
space-averaged flux.

Thus, the transport rate defined derives only from drift and
not from diffusion. Interestingly though, diffusion is indispensable for the mechanism in that in the \textsc{off} state it evolves the concentration distribution,
which affects the above integral once the potential turns back on. More physically, diffusion is what is rectified by the spatial asymmetry to produce directed transport.

\section{Transient Dynamics}

Upon initiating pumping, the dynamics is aperiodic, but it asymptotically
converges to the period of the switching. First we illustrate this behavior, and later provide a physical reasoning.

In Fig. \ref{fig:Cumulative-flux-through}
we plot the numerical time-integral of the transport rate $T(t)$,
i.e., the cumulative number of particles transported over time. The
cumulative local flux, i.e., the time-integral of $J(x,t)$, at two arbitrary positions, is plotted alongside for illustration. These integrals
are reset to zero at the start of each \textsc{on} phase, to aid comparison
between cycles.

In accordance with the previous arguments, the cumulative transport curve is observed to be flat during the \textsc{off}
phases, although the local fluxes vary smoothly over the entire cycle,
and differently at different positions. Interestingly, the dynamics
passes through an intermediate maximum of transport before attaining
periodicity. The entire dynamics attains
periodicity, which is why, for example, the curves of the two local
fluxes also exhibit periodicity at the end. Also interesting is that
in the final periodic dynamics, the transport at the start of each
cycle is briefly backwards (negative) before it compensates and moves forward.

We seek a measure of convergence to periodicity. Call $J_{cyc}(x)$ the integral of $J(x,t)$ over an on-off cycle,
i.e., the number of particles passing through a point over a cycle.
As the dynamics converges to periodicity, one must expect this
to converge to the same value at every position (equal to its
spatial mean $T_{cyc}$, the transport integrated over a cycle). Otherwise
there would be accumulations or depletions between different positions
in each cycle, and they would grow unboundedly now that the dynamics
is periodic, breaking particle conservation.

Thus the standard deviation of $J_{cyc}(x)$ over the spatial period provides such a measure. Dividing by the mean $T_{cyc}$ yields the spatial coefficient of variation, a more uniform measure that accounts for the variation of the asymptotic transport rate itself. We use this as the measure of approach to periodicity, and plot it against cycles on the secondary axis of Fig. \ref{fig:Cumulative-flux-through}. As expected, it is seen to decay asymptotically as the dynamics converges to periodicity.

If such a pumping system operates for a long time, the long-term transport rate shall be the asymptotic periodic one. Thus, in discussions of the transport rate of the model for a choice
of parameters, the quantity of interest is the asymptotic rate, which as we noted is
unfortunately lower than the intermediate maximum.

Now we attempt to explain the transient and long-time periodic behavior via the following mechanism. In each \textsc{on} phase the drift
pushes the concentration towards the potential valleys against a gradually
mounting diffusive resistance. In the \textsc{off} phase the diffusion
undoes some of this by spreading it back out. However, with continuing cycles, the drift succeeds in incrementally advancing this turning point, until the progress due to drift in the \textsc{on}
phase is exactly pushed back by the diffusion in the \textsc{off}
phase. The concentration then settles into a periodic oscillation between the distributions at the turning points, which shall be discussed in further detail.

\begin{figure*}[t]

\begin{minipage}[t]{0.5\textwidth}%
\noindent \begin{flushleft}
\textbf{a}
\par\end{flushleft}
\includegraphics[width=1\linewidth]{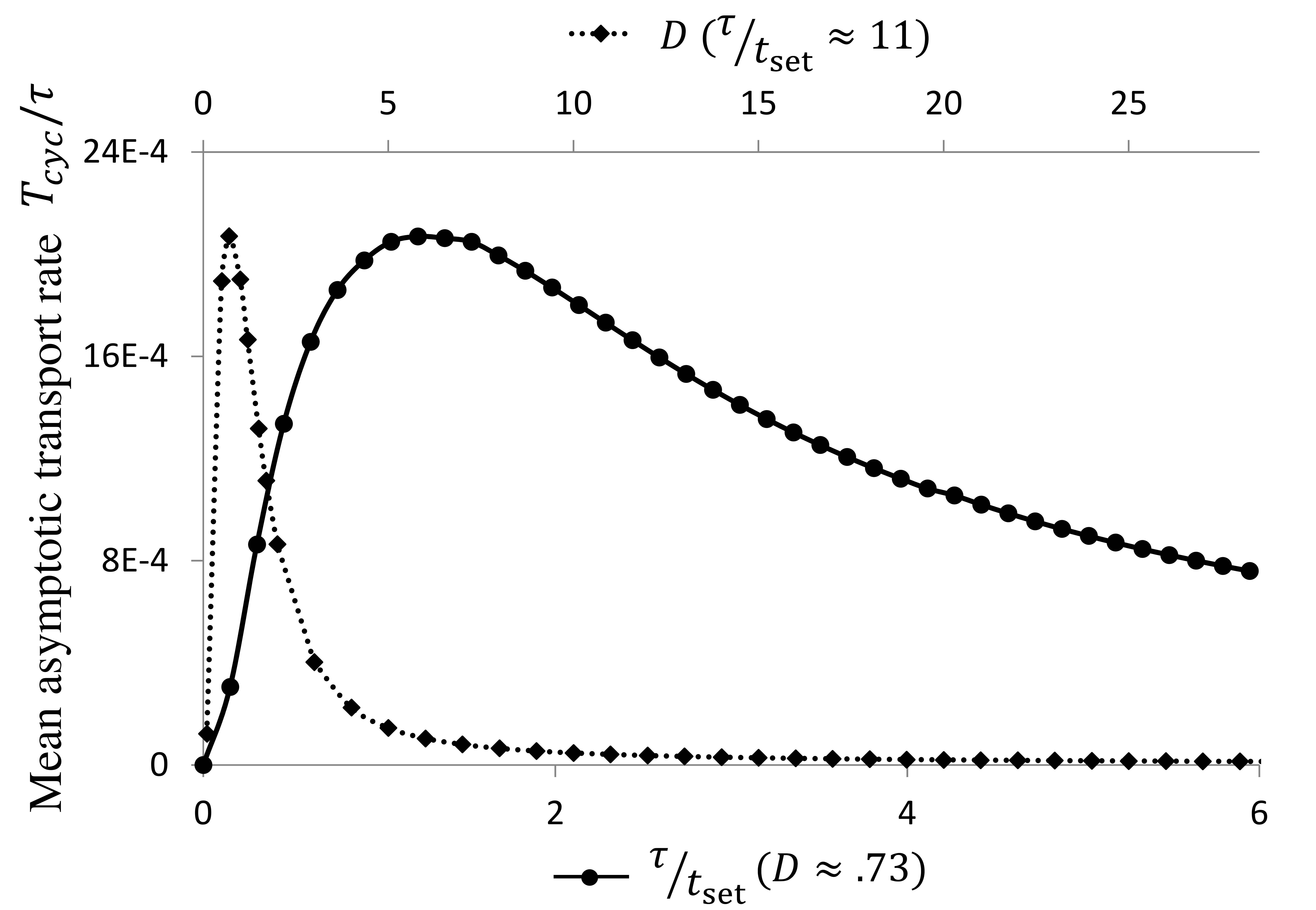}
\end{minipage}\hfill{}%
\begin{minipage}[t]{0.5\textwidth}%
\noindent \begin{flushleft}
\textbf{b}
\par\end{flushleft}

\includegraphics[width=1\linewidth]{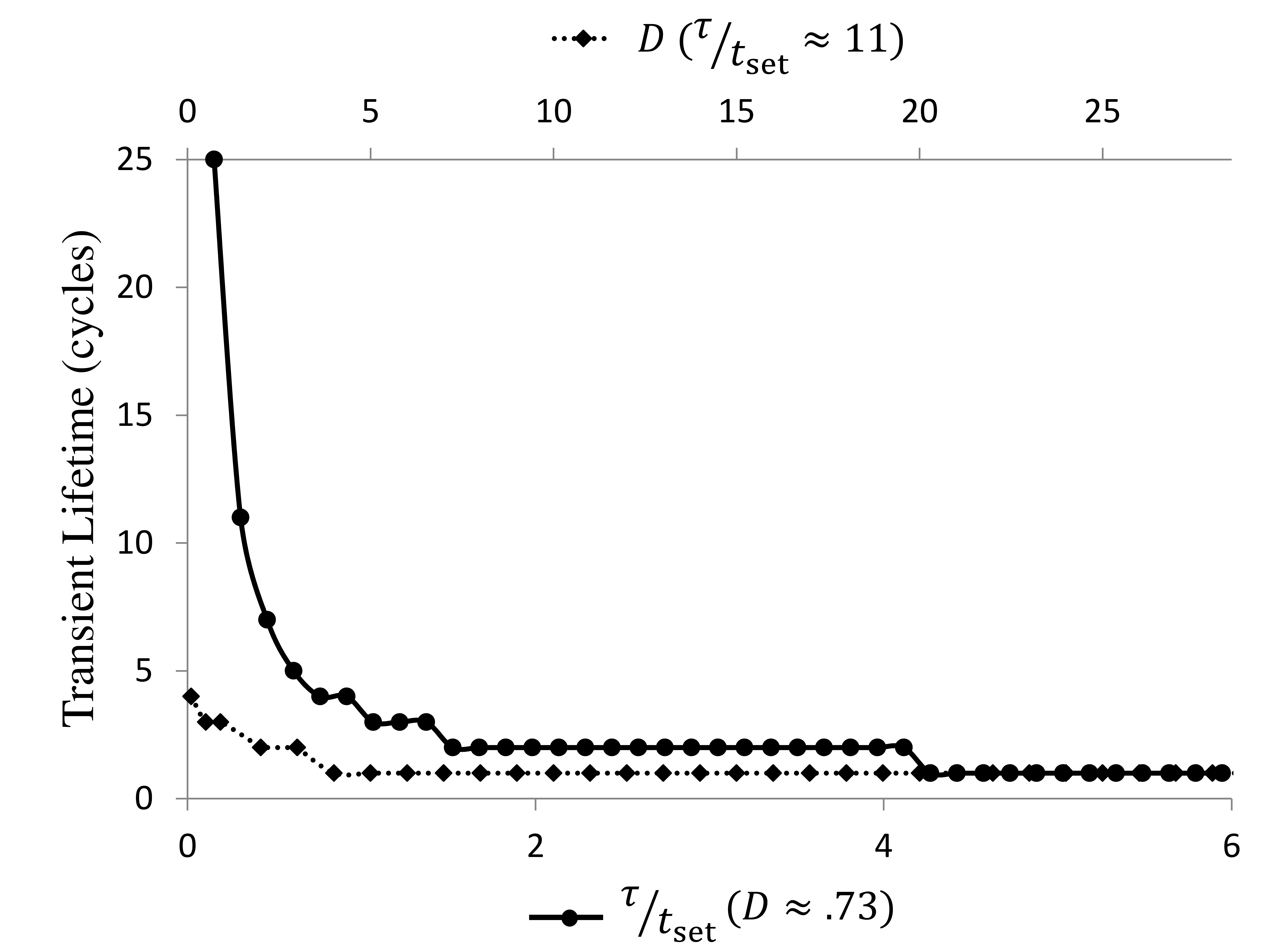}
\end{minipage}
\caption{Effects of varying $\tau$ (primary axis)
and $D$ (secondary axis), at $\alpha=\frac{1}{4}$. Each abscissa
indicates the same value of $\hat{D}$ on both horizontal axes, ranging
from 0 to $\approx.36$ over the horizontal range. $\hat{U}_{\text{max}}$
ranges from 0 to 146 over the $\tau$-axis and is fixed at 30 over
the $D$-axis.
(a) Avg. asymptotic transport rate $T_{cyc}/\tau$ (normalized
by initial concentration height) exhibits a local maximum against
both $D$ and $\tau$. \label{fig:A:-Asymptotic-transport}
(b) Number of cycles required to attain periodicity drops
asymptotically with both increasing $D$ and $\tau$.}

\end{figure*}

\section{Transient Lifetime and Long-term Transport Rate}

In light of the previous discussion, we first define the transient
lifetime to be the number of cycles elapsed until the coefficient
of variation of $J_{cyc}(x)$ drops below 1\%. We observed that both
the transient lifetime, and the average asymptotic rate $T_{cyc}/\tau$
that the transport converges to, depend on the diffusion coefficient $D$ as well as the duration
of the switching cycle $\tau$. To study the latter, we use the `settling
time' as a time scale of the system against which to measure the switching
period. This is defined as the maximum time required to equilibriate
in the \textsc{on} phase when we disregard diffusion.
It is given by the time taken by a particle to traverse the gentler ratchet
slope with the drift speed in that region, and is given by $t_{\text{set}}=b^{2}\gamma/U_{\text{max}}$.

In Fig. \ref{fig:A:-Asymptotic-transport} (a) and (b)
we plot the variation of the transient lifetime and the asymptotic
average transport rate (normalized by the uniform starting
concentration height) in response to varying either $\tau$ or $D$ while keeping the other fixed. In the dimensionless formulation, both amount to varying $\hat{D}$ proportionally. Therefore,
in plots (a) - (c) of Fig. \ref{fig:A:-Asymptotic-transport},
the scales of the two horizontal axes have been adjusted so that any
abscissa indicates the same value of $\hat{D}$ on either axis, which
ranges from 0 to $\approx.36$ over the horizontal range of the plot.
The two curves differ since varying $\tau$ at fixed $D$ also
entails varying $\hat{U}_{\text{max}}$ proportionally (it ranges
from 0 to $\approx146$ over the $\tau$-axis), while upon varying
$D$ at fixed $\tau$ it is unchanged (fixed at 30 over the $D$-axis).
Yet, their trends are similar, due to the following reason.

The asymptotic transport rate exhibits a local maximum against both
$\tau$ and $D$, i.e., it resonates at a particular combination of
forcing frequency and diffusivity, reproducing a previous result by
\cite{1994Prost}. It vanishes at large values of both for different
reasons. For large $\tau$ the dynamics approaches steady state at
the end of the switching times, and particles stop being transported.
The transport rate per unit time thus falls against increasing $\tau$.
At large $D$ the concentration hardly responds to the potential,
barely moving from the initial flat profile, impeding transport.

Transient lifetime drops with increasing period length and diffusivity.
Increasing $D$ at fixed $\tau$ causes the incremental advance with continuing cycles due
to drift to stop sooner, while increasing $\tau$ at fixed $D$ allows the
drift more time to advance the concentration until it rests against
diffusive pressure. Both lead to the attainment of the periodic oscillatory
dynamics within fewer cycles.

\begin{figure}
\includegraphics[width=1\linewidth]{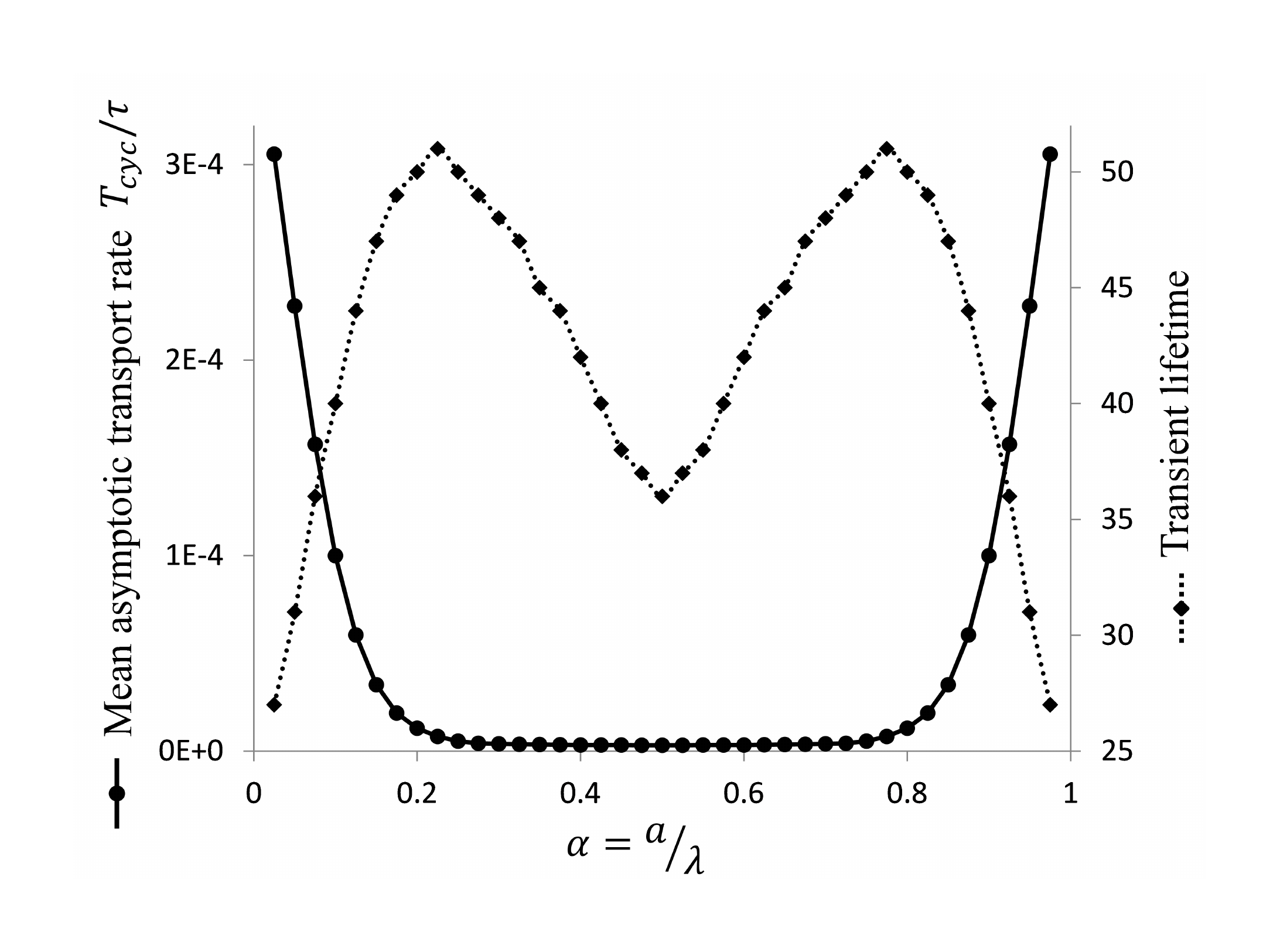}
\caption{Responses of the mean (absolute) asymptotic transport rate and the transient lifetime, with
varying skewness $\alpha$ of the ratchet shape. Parameters: $\hat{D}=.0025, \hat{U}_{\text{max}}\approx.023.$\label{fig:transportvsa}}
\end{figure}

The responses of the mean asymptotic transport rate and the transient lifetime to the degree of skewness
of the ratchet have been illustrated in Fig. \ref{fig:transportvsa}. The behavior of the transport rate is more readily understood. $\alpha\lessgtr\frac{1}{2}$
corresponds to the right (left) slope being the gentler slope, causing transport in that direction. (We have plotted the absolute rate, i.e. positive in both directions.) The greater the skewness, the higher the rectification and therefore the transport rate. A symmetric ratchet shape naturally yields no transport.

The explanation behind the variation of the transient lifetime is more involved. As described previously, periodicity is attained when a balance is struck between the advance of the particles by drift and their retrograde diffusion over each cycle. The relative strengths of the two determine how many cycles elapse until this balance is attained. High and low skewness tilt the balance strongly in favour of the drift and diffusion respectively, both assisting in quicker attainment of the final oscillatory equilibrium. An intermediate skewness causes the concentration to both advance and roll back by a great extent in each cycle, which takes longer to settle into an exactly balanced rhythm.

\section{Asymptotic Periodic Oscillation between Turning Curves}

\begin{figure*}

\begin{minipage}[t]{0.5\textwidth}%
\noindent \begin{flushleft}
\textbf{a}
\par\end{flushleft}
\includegraphics[width=1\linewidth]{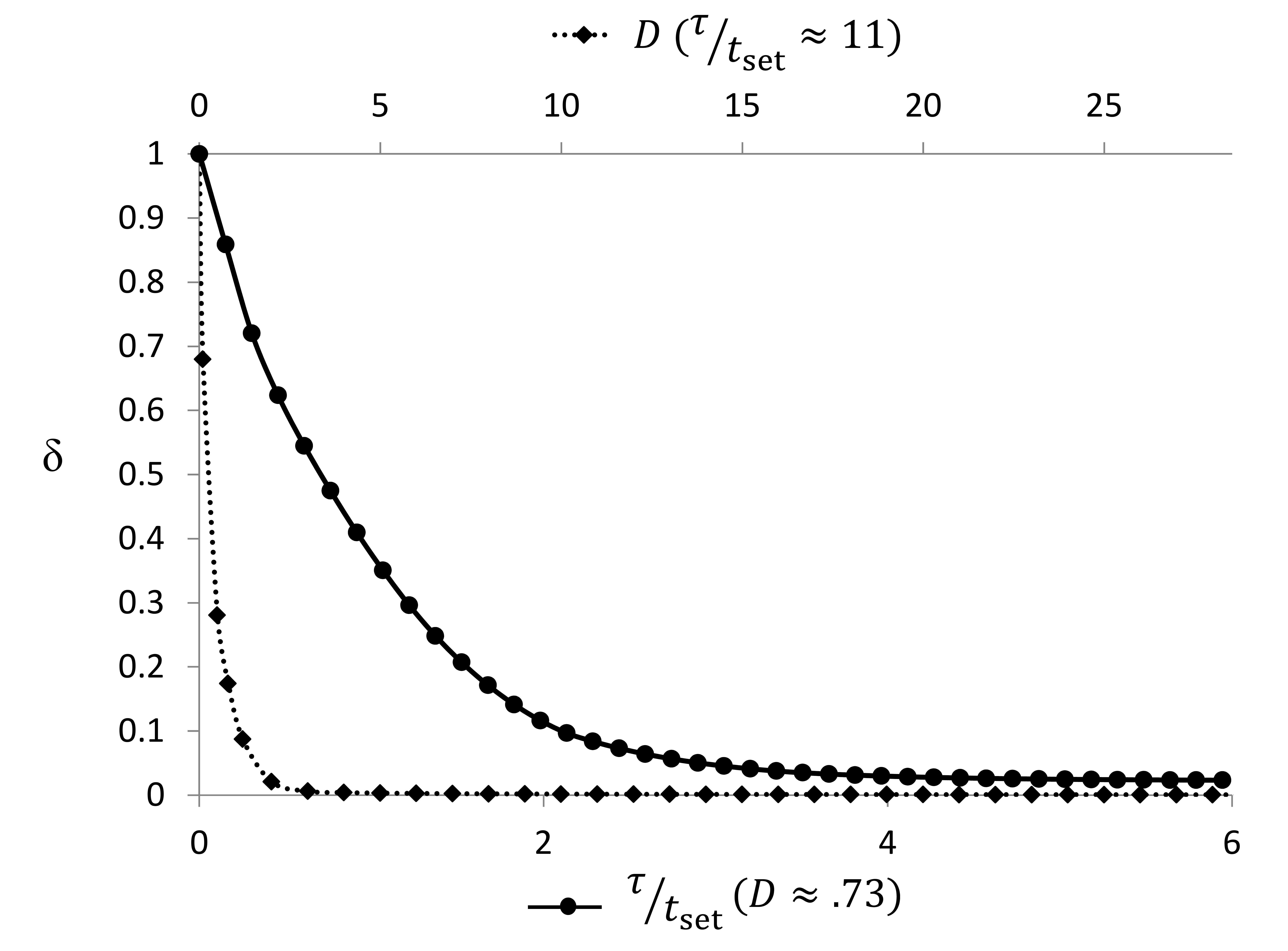}
\end{minipage}\hfill{}%
\begin{minipage}[t]{0.5\textwidth}%
\noindent \begin{flushleft}
\textbf{b}
\par\end{flushleft}
\includegraphics[width=1\linewidth,trim=0 0 1 1,clip]{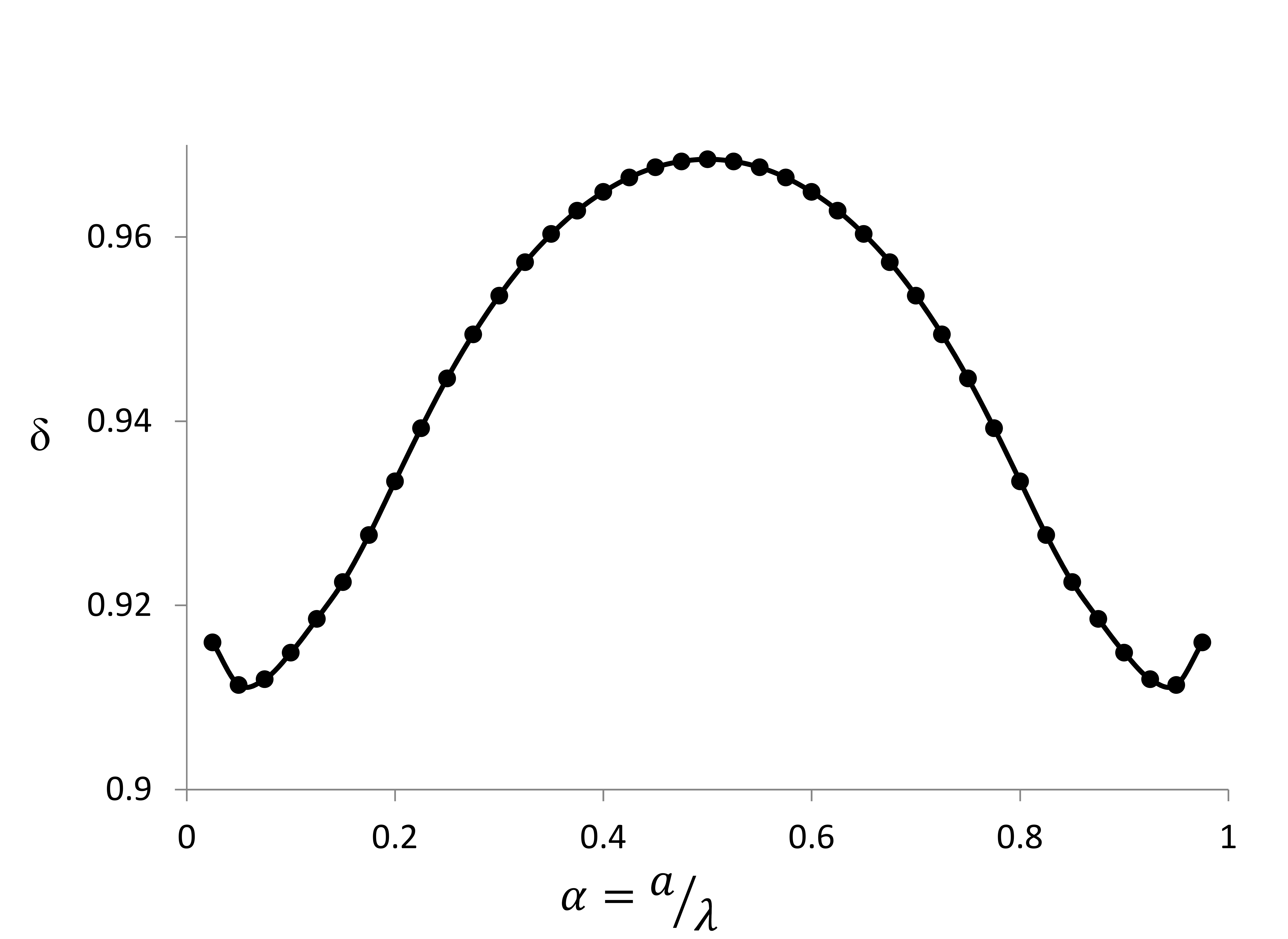}
\end{minipage}
\caption{Response of $\delta$, a measure of deviation of the turning
curves from the steady state distributions, against some system parameters.
(a) $\delta$ decays with increasing $D$ and $\tau$. Parameter values and axis scale adjustments same as Fig.
\ref{fig:A:-Asymptotic-transport}. \label{fig:L1vsperiod}
(b) $\delta$ exhibits a maximum for a symmetric ratchet potential shape and a minimum for an intermediate degree of skewness $\alpha$. However, the overall variation is small. Parameters same as Fig. \ref{fig:transportvsa}.}

\end{figure*}

The steady state Boltzmann distribution consists of two piecewise exponentials in the \textsc{on} phase of the potential, and is flat for the \textsc{off}
phase (dashed curves in Fig. \ref{fig:turning-curves}). Call these
$c_{\text{eq}}^{\text{on}}(x)$ and $c_{\text{eq}}^{\text{off}}(x)$
respectively. If the concentration reaches steady state at the
end of either or both phases, the dynamics
shall be identical in each cycle. For example, if the concentration reaches equilibrium at the end of each
\textsc{on} phase,
the diffusive dynamics of the \textsc{off} phase shall begin from
this steady state and proceed identically in each cycle until it is interrupted
at the same point by the potential turning \textsc{on}.

However, switching the potential with a finite period prevents the concentration from reaching equilibrium in either phase. In the asymptotic behavior, the dynamics is nevertheless periodic, as the concentration on its way to either of the equilibria is interrupted
at the same two points by the potential switching.
The solid curves in Fig. \ref{fig:turning-curves} represent these
`turning point' profiles at the switchings, between which the
concentration oscillates. Call the turning curves at the end of the
\textsc{on} and \textsc{off} phases as $c_{\text{turn}}^{\text{on}}(x)$
and $c_{\text{turn}}^{\text{off}}(x)$ respectively.

Oscillation between the two steady states would allow the maximum possible `amplitude', but is not possible with a finite switching period. The turning curves represent the end points of a restricted, smaller amplitude oscillation. A natural quantity to study therefore is the departure of the turning curves from the corresponding steady state distributions, representing the fraction of the maximum amplitude covered. To quantify this, we measure the difference
in shape between two curves by the $L_{1}$-distance within a period:
$d_{1}(f,g)=\int_{\lambda}\left|f(x)-g(x)\right|dx$, i.e. the unsigned
area contained between them (highlighted in Fig. \ref{fig:turning-curves}). We consider $d_{1}(c_{\text{eq}}^{\text{on}},c_{\text{turn}}^{\text{on}})+d_{1}(c_{\text{eq}}^{\text{off}},c_{\text{turn}}^{\text{off}})$
as a measure of departure of the turning curves from the steady states. We normalize this as a fraction of the distance between the two steady states $d_{1}(c_{\text{eq}}^{\text{on}},c_{\text{eq}}^{\text{off}})$ and define:
\[
\delta=\frac{d_{1}(c_{\text{eq}}^{\text{on}},c_{\text{turn}}^{\text{on}})+d_{1}(c_{\text{eq}}^{\text{off}},c_{\text{turn}}^{\text{off}})}{d_{1}(c_{\text{eq}}^{\text{on}},c_{\text{eq}}^{\text{off}})}
\]

We plot $\delta$ in Fig. \ref{fig:L1vsperiod} (a) against increasing switching period (measured in units of $t_{\text{set}}$), and $D$, each while keeping the other constant as before.

Increasing the forcing period allows the concentration more relaxation
time to evolve towards steady state in each phase, and the turning
profiles should be closer to the corresponding Boltzmann equilibrium profiles. 
Accordingly, $\delta$ was seen to decrease smoothly and monotonically with increasing switching period and go asymptotically to 0 for long periods.
It is greatest in the limit of zero switching period, when the concentration
does not move from its initial uniform distribution (which is also
$c_{\text{eq}}^{\text{off}}$), and the ratio is therefore 1.

The response to varying $D$ is similar, but its explanation is different. At the limit of vanishing $D$, diffusion barely rolls back the advance due to drift in each cycle, and thus with continuing cycles the concentration incrementally and asymptotically advances towards the \textsc{on} phase steady state $c_{\text{eq}}^{\text{on}}$, and is hardly pushed back in the \textsc{off} phase. Therefore, in the asymptotic periodic dynamics, the profiles $c_{\text{turn}}^{\text{on}}$ and $c_{\text{turn}}^{\text{off}}$ are both almost the same as $c_{\text{eq}}^{\text{on}}$. $c_{\text{eq}}^{\text{off}}$, however, is still the flat profile (at $D=0$, $c_{\text{eq}}^{\text{off}}$ would also be the same as $c_{\text{eq}}^{\text{on}}$, but recall that we are instead considering $D\rightarrow0$). Therefore, both the numerator and denominator of $\delta$ become $d_{1}(c_{\text{eq}}^{\text{on}},c_{\text{eq}}^{\text{off}})$ in this limit, yielding 1.

At high $D$, $c_{\text{eq}}^{\text{on}}$ is only slightly deviated from the flat $c_{\text{eq}}^{\text{off}}$, yielding a small denominator. As noted before, the approach towards the steady state $c_{\text{eq}}^{\text{on}}$ is faster at higher $D$, so that $c_{\text{turn}}^{\text{on}}$ is almost coincident with it. Also the high diffusion rapidly rolls back this drift almost entirely in every \textsc{off} stage, so that $c_{\text{turn}}^{\text{off}}$ is also nearly coincident with $c_{\text{eq}}^{\text{off}}$. Thus, with rising $D$, both terms in the numerator drop more sharply than the denominator, yielding a decaying $\delta$.

As the skewness $\alpha$ of the ratchet potential shape is varied, $\delta$ exhibits a maximum for a symmetric ratchet shape, and a minimum for an intermediate skewness, but the variation is overall weak compared to the response against other parameters. This variation is related to details of how the shapes of the equilibrium and turning curves shift in response to changing skewness, and does not have a strong and immediate physical cause.

\section{Conclusions}

We employed a numerical solution of the governing drift-diffusion
equations of an infinite 1D thermal ratchets model to yield its full dynamics, and observed that initially aperiodic dynamics converges to the period of the switching.
We defined measures of the large-scale transport achieved by the system and the transient lifetime. The
transport rate maximizes at a finite diffusivity $D$ and switching
period $\tau$, and increases with greater skewness of the asymmetric
ratchet shape. The transient lasts for fewer switching cycles upon increasing
$D$ or $\tau$, and lasts longest at an intermediate skewness.

The long-term periodic dynamics of the concentration
oscillates between turning curves, and we defined a measure of their deviation from the equilibrium distributions that corresponds to the extent of the maximum `amplitude' covered. This deviation asymptotically decays for high $D$ and $\tau$, and is highest for an intermediate skewness, although this last variation is weak.

Further work could involve revisiting previous studies of separating particles with different diffusivities (see \cite{2002Reimann}) in the light of these dynamical results. Another suggestion is to lift the restriction of equal \textsc{on} and \textsc{off} durations and observing the effect of their ratio on dynamical features. The constraint of spatial periodicity in the dynamics may also be removed. This will allow studying the nature of the spatial aperiodicity and whether and how convergence to spatial periodicity (i.e. long-range order) occurs, analogous to the temporal observations carried out here.

\bibliographystyle{apsrev4-1}
\nocite{*}
\bibliography{paper}

\end{document}